\documentclass{article}
\usepackage[T1]{fontenc}
\usepackage[utf8]{inputenc}
\usepackage[lbd]{ismir}
\usepackage{amsmath,cite,url}
\usepackage{graphicx}
\usepackage{color}
\usepackage{microtype}

\title{ScorePrompts: Natural-Language Exploration \\ of Symbolic Music Scores through Analysis}

\oneauthor
  {Emmanouil Karystinaios \hspace{1.2cm} Gerhard Widmer}
  {Institute of Computational Perception, Johannes Kepler University Linz, Austria\\
   {\tt emmanouil.karystinaios@jku.at}}

\def\authorname{E. Karystinaios, G. Widmer}
\usepackage[bookmarks=false,
  pdftitle={ScorePrompts: Evidence-Linked Natural-Language Interaction with Symbolic Music Scores},
  pdfauthor={\authorname},pdfsubject={\pdfsubject},hidelinks]{hyperref}

\begin{document}

\maketitle

\begin{abstract}
We present ScorePrompts, an interactive system in which users upload a score, receive natural-language descriptions of its musical structure, ask questions about specific passages, and inspect the corresponding analysis results in staff notation. Specialist MIR components first estimate harmony, tonality, cadences, formal boundaries, texture, and note-level roles, organizing their outputs at note, beat, measure, and piece levels. A schema-constrained language model converts these results into descriptions rather than inferring musical structure directly from raw MusicXML. For questions such as “What changes in measures 14–18?”, a deterministic router selects the relevant measures and analytical levels and returns a concise response together with the underlying results and caveats. Verovio renders the score and links the returned information to cited measures and note-level attributes. The interface also exposes intermediate tables and disagreements between analytical levels. ScorePrompts is intended for exploratory score analysis and explanation, not score editing. The demo shows how existing analysis models, constrained language generation, Q\&A retrieval, and notation-based visualization can provide natural-language access to symbolic music analysis while keeping intermediate results inspectable.
\end{abstract}

\section{Introduction}

Graphical notation encodes pitch spelling, simultaneity, voices, meter, and large-scale organization in a representation designed for musical reading. These relations are easily obscured when a score is flattened into a token sequence or exposed as raw MusicXML. Natural language can provide a useful point of entry, but language alone is insufficient: a user should be able to inspect which score regions and analytical layers support an explanation.

ScorePrompts follows an \emph{analyze-before-generating} design. Rather than asking a language model to infer musical structure directly from markup, the system first constructs a typed, score-aligned representation using MIR models and deterministic descriptors. Language generation and score questions operate on this representation, while the intermediate analyses remain visible. The contribution is a modular interaction workflow combining multi-resolution evidence alignment, schema-constrained explanation, deterministic query-directed retrieval, and notation-linked inspection. A public web demo\footnote{\url{hf.co/spaces/manoskary/scoreprompts}} exposes the complete workflow.

\section{Related Work}

Language access to symbolic scores has been studied through passage retrieval, symbolic-language modeling, tool-augmented reasoning, captioning, and pedagogical dialogue. C@merata relates English descriptions to passages in MusicXML scores~\cite{sutcliffe2015relating}, while ChatMusician treats ABC notation as a language-modeling medium~\cite{yuan2024chatmusician}. Some benchmarks further show that multi-level musical reasoning remains difficult for general language models~\cite{dai2026msubench}.

Several recent systems are close to the present setting. CSyMR combines language-model control with deterministic \texttt{music21} operators for compositional retrieval~\cite{wang2026csymr}. MIDI-PHOR distills structured MIDI views for evidence-linked captioning~\cite{au2026midiphor}, and MelodyMate combines ABC-grounded dialogue with graphical and audio views for music-form learning~\cite{jin2026melodymate}. MuseAgent-1 integrates OMR, performance-audio analysis, and agentic multimodal reasoning over structured representations~\cite{zhao2026museagent}. ScorePrompts instead assumes machine-readable MusicXML and places a fixed, inspectable MIR stack before language generation. 

\begin{figure*}[t]
  \centering
  \begin{minipage}[t]{0.55\textwidth}
    \centering
    \includegraphics[
      alt={ScorePrompts question view. A question about a measure is answered with a takeaway, measure, beat, and note evidence, caveats, and an evidence table.},
      viewport=0 169.5 856.5 627,clip,width=\linewidth
    ]{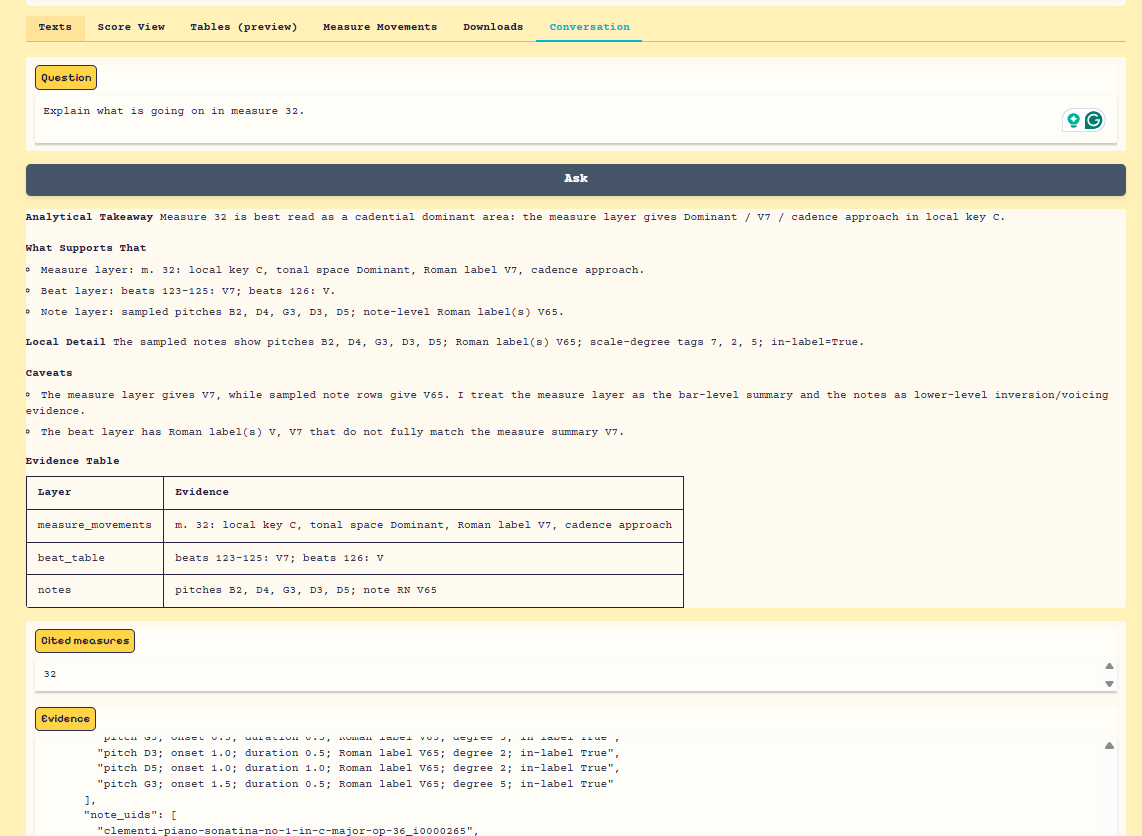}\\
    \textbf{(a)} Multi-resolution evidence for a score question
  \end{minipage}\hfill
  \begin{minipage}[t]{0.365\textwidth}
    \centering
    \includegraphics[
      alt={ScorePrompts Verovio view showing a piano score with analytical cues for score and note inspection.},
      viewport=0 380 1050 933,clip,width=\linewidth
    ]{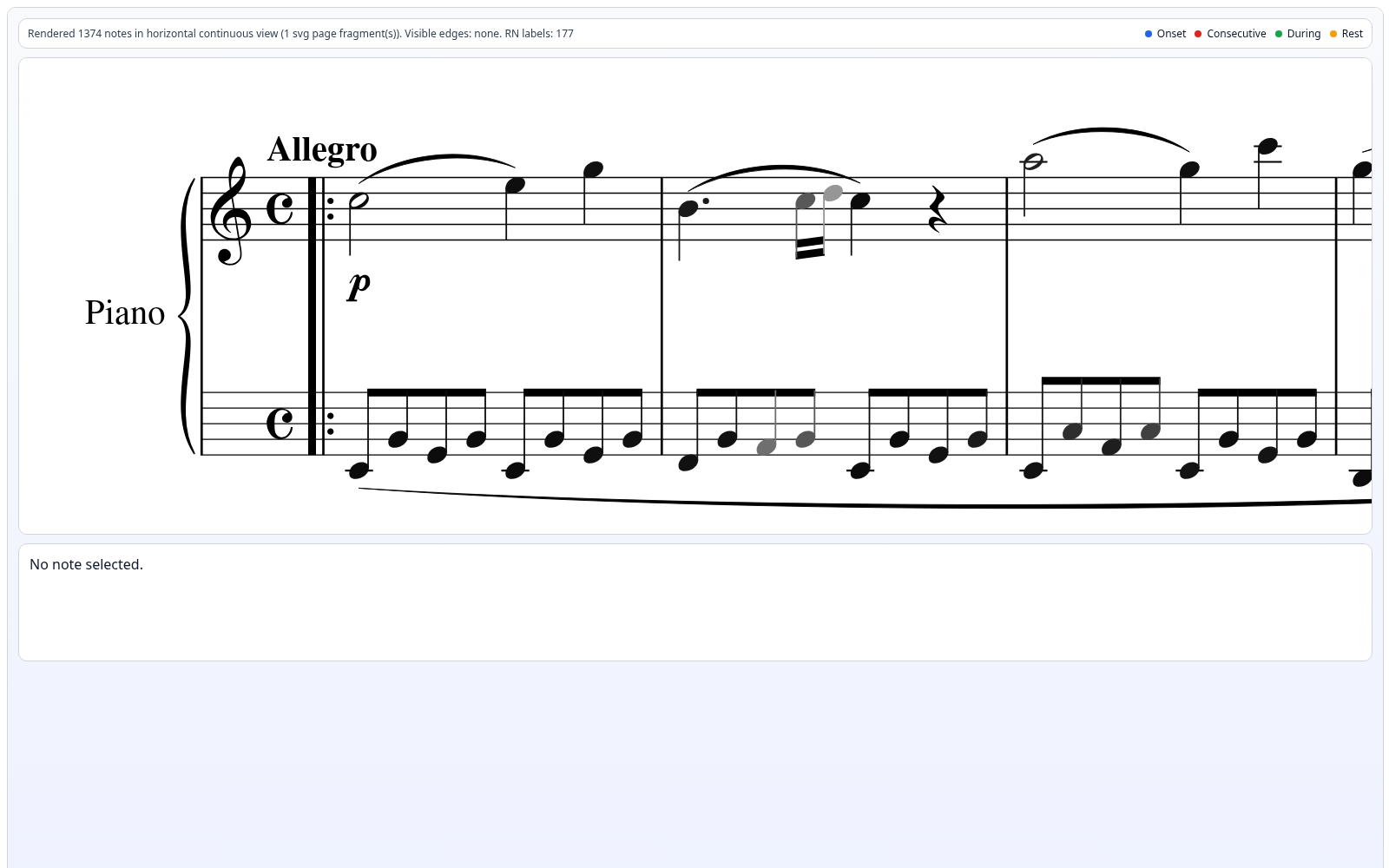}\\
    \textbf{(b)} Notation-linked inspection
  \end{minipage}
  \caption{Representative interface views from separate runs. (a) Deterministic retrieval returns measure-, beat-, and note-level analysis results together with caveats and cross-level disagreement. (b) Verovio links the analytical representation to staff notation and note inspection.}
  \label{fig:interface}
\end{figure*}

\section{System Design}

\subsection{Specialist Analysis and Evidence Alignment}

Partitura~\cite{cancino2022partitura} ingest an MusicXML score and loads score-aligned events, and AnalysisGNN~\cite{karystinaios2025analysisgnn} estimates harmonic, tonal, cadential, phrase, section, and note-role attributes. Deterministic post-processing reconstructs Roman numerals and aggregates predictions at beat and measure levels. AlgoMus-derived texture descriptors~\cite{couturier2022texture}, jSymbolic-style features~\cite{mckay2018jsymbolic} computed with \texttt{music21}~\cite{cuthbert2010music21}, and score metadata provide complementary context.

The resulting note-, beat-, measure-, and piece-level views are aligned through measure references and pipeline-assigned note identifiers. The complete score is represented as a sequence of overlapping measure chunks carrying field definitions, categorical codebooks, and local analysis results rows. Chunking is a transport mechanism for bounded model contexts; a score-level evidence index preserves access to all decoded chunks and to global trajectories in key, tonal space, cadence, stability, and section.

\subsection{Constrained Language Generation}

Symbolic notation is not sent to the language model. Instead, schema-constrained stages transform analytical rows into canonical measure facts, organize those facts together with piece-level trajectories, and realize three descriptions with increasing technical specificity. Each chunk can be compiled independently and its facts incorporated into a score-level plan. References produced by the language stages are checked against the evidence index, and malformed structured outputs are retried or replaced by deterministic summaries. The language component is model-agnostic and uses no music-specific fine-tuning; the scientific object passed to it is the typed MIR representation rather than the score markup itself.

\subsection{Deterministic Score Questions}

Supported score questions bypass an additional and expensive language-model call. A deterministic router parses explicit measures and ranges together with terms concerning harmony, cadence, form, and analytical results. It then selects the relevant fields from the complete chunk index, using measure-level trajectories as the primary context and adding beat- or note-level rows when required. The response separates a concise takeaway from cited measures, supporting layers, caveats, and structured analysis references, as illustrated in Figure~\ref{fig:interface}(a). Exposing the selected rows makes disagreements between analytical resolutions visible rather than hiding them behind a single verbal answer.

\subsection{Notation-Linked Inspection}

The original notation is retained in parallel with the analytical representation. Verovio~\cite{pugin2014verovio} renders the score with reconstructed Roman-numeral labels. Selecting a note reveals the same identifiers used by the evidence index, together with pitch, timing, harmonic and structural labels, and available confidences. This closes the interaction loop by returning natural-language explanations to the score regions from which their analysis was derived.

\section{Interactive Demonstration}

The demo follows one score through the four stages of the system. After analysis, the user can compare the descriptions with the underlying note-, beat-, measure-, and piece-level tables. A focused question, such as ``What changes in measures 14--18?'', shows how the router narrows the analysis results and reports missing or conflicting information. The cited measures can then be inspected in the rendered notation, and individual notes can be selected to show their local analytical attributes. Complete note, beat, and measure tables are downloadable as CSV, together with a global JSON summary. The sequence is designed to make both successful explanations and questionable model outputs inspectable under the same analytical stack.

\section{Discussion and Scope}

ScorePrompts makes analysis outputs the primary evidence layer rather than asking a language model to replace symbolic analysis. The separation of analysis, language generation, retrieval, and rendering supports independency and evaluation of each component. The analysis results can be traced to score-derived fields and valid measure or note identifiers, however, we do not establish that each generated sentence is semantically linked to those fields, nor that the upstream predictions are musicologically correct.

The current prototype targets machine-readable Western tonal scores and inherits the repertory and annotation assumptions of its MIR components. The representation and generation stages are defined per chunk and can be aggregated over the complete score. The present system does not edit notation and has not yet been evaluated in a user study. Natural next steps are claim faithfulness checks, full-score description aggregation, RAG-augmented retrieval for Q\&A and task-based evaluation with musicians.

\section{AI Usage Statement}

Generative AI tools were used for code assistance, literature discovery, and language editing. The authors verified all claims, citations, and final text.

\section{Acknowledgements}
This research has been supported by the European Research Council (ERC) under the EU's Horizon 2020 research \& innovation programme, grant agreement No. 101019375 (Whither Music?).

\bibliography{references}

\end{document}